\documentclass[preprint,showpacs,preprintnumbers,amsmath,amssymb]{revtex4}
\usepackage{amsfonts}
\usepackage{amssymb}
\usepackage{latexsym}
\usepackage{psfig}
\usepackage{graphicx}

\newcommand{\al}{\alpha}

\newcommand{\X}{{\cal X}}

\newcommand{\rar}{\rightarrow}

\begin{document}

\begin{flushright}
 Preprint ICN-UNAM 16/04
\end{flushright}

\title{Hydrogen atom and one-electron molecular systems in a
strong magnetic field: are all of them alike?}

\author{A.~V.~Turbiner}
\email{turbiner@nuclecu.unam.mx}
 \altaffiliation[]{On leave of absence from the ITEP,
 Moscow 117259, Russia}

\author{A.~B.~Kaidalov}
\email{kaidalov@heron.itep.ru}
 \altaffiliation{Institute for Theoretical
 and Experimental Physics (ITEP), Moscow 117259, Russia}

\author{J.~C.~L\'opez Vieyra}
\email{vieyra@nuclecu.unam.mx}
\affiliation{%
Instituto de Ciencias Nucleares, Universidad Nacional Aut\'onoma
de M\'exico, Apartado Postal 70-543, 04510 M\'exico, D.F.,
M\'exico}

\date{\today}

\begin{abstract}
  Easy physics-inspired approximations of the total and
  binding energies for the ${\rm H}$ atom and for the molecular ions
  $${\rm H}_2^{(+)}\, ({\rm ppe}),\, {\rm H}_3^{(2+)}\, ({\rm pppe}),\,
  ({\rm HeH})^{++}\, (\al {\rm p e}),\, {\rm He}_2^{(3+)}\, (\al \al {\rm e})$$
  as well as quadrupole moment for the ${\rm H}$ atom and the equilibrium
  distances of the molecular ions in strong magnetic fields $> 10^{9}$\,G
  are proposed. The idea of approximation is based on the assumption that
  the dynamics of the one-electron Coulomb system in a strong magnetic
  field is governed by the ratio of transverse to longitudinal sizes of
  the electronic cloud.
\end{abstract}

\pacs{31.15.Pf,31.10.+z,32.60.+i,97.10.Ld}

\vskip 2cm

\begin{center}
 {\it Invited contribution to Czechoslovak Chemical
 Communications \\ to a Special Issue in honor of Professor
 Josef Paldus}
\end{center}

\maketitle


The behavior of Coulomb systems in a strong magnetic field has always
attracted a lot of attention. This was justified by the presence of
strong magnetic fields in astrophysics (neutron
stars and white dwarfs \footnote{Values of the observed magnetic
fields are: white dwarfs ($10^6-10^9$\,G), magnetic neutron stars
($10^{12}-10^{13}$\,G), magnetars ($10^{14}-10^{15}$\,G)}), as well as in
plasma and semiconductor physics. In particular, for many years it
has existed a question about the content of the neutron star
atmosphere. Since the seminal papers by Kadomtsev-Kudriavtsev
\cite{Kadomtsev:1971} and Ruderman \cite{Ruderman:1971} it was
believed that the neutron star atmosphere subject to a strong
magnetic field is made from atomic-molecular compounds. However,
in order to construct a model of the atmosphere it is necessary to
explore matter and its properties in a strong magnetic field.

Recently, it was discovered that the interplay of Coulomb and
magnetic forces for $B \gtrsim 10^{11}$\,G leads to a new physics:
new bound one-electron Coulomb systems appear like the exotic
molecular ions ${\rm H}_3^{++}$ \cite{Turbiner:1999}, $({\rm HeH})^{++}$ and
${\rm He}_2^{3+}$ \cite{Turbiner:2004He}. These Coulomb systems do not
exist without magnetic field. All of them are characterized by
very large binding energies growing with magnetic field. For all
magnetic fields, where the non-relativistic considerations are
justified ($B < 4.414 \times 10^{13}$\,G), of one-electron
atomic-molecular systems the hydrogen atom - the only neutral
system- is characterized by the highest (!) total energy, being
correspondingly the least bound system \cite{Lopez-Tur:2000}. At
the same time it seems natural to assume that
more-than-one-electron Coulomb systems in a strong magnetic field
are not strongly bound (if bound) due to the fact that all
electron spins should be parallel, being antiparallel to the
magnetic field direction.

All one-electron molecular systems have a certain common feature.
Their optimal configuration is always a configuration where all
massive charged centers are situated on a magnetic line. We call
it {\it parallel configuration}. Only these configurations are
considered in the present article.

It is well known that studies in a strong magnetic field are very
complicated for several reasons. Perhaps, the most serious,
conceptual reason is related to the fact that the bound states are
of a weakly-bound-state nature (the binding energies are much
smaller than the total ones). The perturbation theory in powers of $B$
is fast divergent and thus cannot be used. Asymptotic expansions
at $B = \infty$ have extremely complicated form but, usually,
have no domain of applicability inside  non-relativistic
considerations. For example, it can be easily checked that at an
extremely strong magnetic field near the edge of applicability of
non-relativistic approximation $B=10000\, {\rm a.u.} (=2.35 \times
10^{13}\,{\rm G})$ \footnote{In dimensionless units (a.u.) the parameter
of expansion is enormous, $10^5$ }, the ionization energy $E_{\rm b}$
calculated numerically differs by 300\% (!) from the value
obtained using the leading term in asymptotic expansion
\[
  E_{\rm b}^{\rm asymp} \ = \ \frac{1}{2}\log^2 B\ ,
\]
where $E_{\rm b}$ is given in atomic units as well as $B$ (see
\cite{Karnakov:2003}). Another example is given by the ratio of
the binding energies of ${\rm H}_2^+$ and ${\rm H}_3^{++}$. Asymptotically, for
$B$ tending to infinity, this ratio should be 1/2.25 . However, the
numerical results at $B=3 \times 10^{13}\,$G \cite{Lopez-Tur:2000}
give $\approx 1$ for this ratio. Therefore the only methods which
can be used are either numerical or variational. For $B \gtrsim
10^{11}$\,G, to the best our knowledge, the numerical methods were
used for the hydrogen atom only (see, for example, the excellent
early review \cite{Garstang:1977}, the book \cite{Ruder} and
recent review articles \cite{Liberman:1995,Lai:2001}). Usually,
these methods are very slow-convergent and extremely difficult to
implement. The most popular method to study one-electron molecular
systems is the variational method (for review, see
\cite{Turbiner:2005}). However, the use of the variational method is
associated with a difficult procedure of minimization and, sometimes,
with numerical calculation of multidimensional integrals with high
accuracy, which can also be quite cumbersome. In any case, the
calculations are made for some particular values of magnetic
field. It seems natural to create some approximate expressions
valid for all magnetic fields, even having not high accuracy, in
order to make at least rough estimates.

The accurate results of calculations of different quantities for
low-lying states reveal a smooth, simple-looking behavior with
rather slow changes with magnetic field. However, a
straightforward attempt to construct approximations either fails
or leads to quite complicated expressions, at least, at first
sight (see, e.g., \cite{Salpeter:1996, Potekhin:2001}). Physical
intuition gives a feeling that there must exist a certain
qualitative technique, for example a type of semi-classical
approximation providing an approximate qualitative description of
these results. So far it is not clear how such a technique can be
approached. A goal of this paper is to consider a certain
simple alternative to this unclear-how-to-approach
technique - to build approximations of the main characteristics of
the one-electron atomic-molecular systems in a constant uniform
magnetic field in their lowest state, such as total and binding
energies, equilibrium distances, electron
cloud sizes, quadrupole moment, by following simple physical arguments.
Our basic assumption is that the physics is mainly governed by a
single parameter: {\it the ratio of the transverse to
longitudinal size of the electron cloud}. Of course, in
dimensionful quantities such as equilibrium distances or
quadrupole moment the transverse and longitudinal sizes should
appear explicitly but only in a form of parameters which carry a
dimension. Hereafter we denote the transverse size of the
electron cloud as $r_{\rm t}$, and the longitudinal size as $r_{\rm l}$.

As always we consider the one-electron Coulomb systems with
infinitely-heavy charged centers, protons and/or $\al$-particles
(the Born-Oppenheimer approximation of the zero order) situated on
the $z$-axis\footnote{It has long been recognized that in a strong
magnetic field ($>10^{12}\,$G) the effects of finite nuclear masses
and center of mass motion are non-trivial. Very few quantitative
studies which exist are very difficult and mostly limited to atomic
type systems. These effects mostly influence the excited states
(for discussion, see \cite{Lai:2001} and references therein).
Our consideration is focused
on the ground states.}. If these charged centers
are of the same charge, they are assumed to be identical.
Although we use the word 'proton' it implies that  in
the Born-Oppenheimer approximation it can be deuteron or triton.
The magnetic field of strength
$B$ is directed along the $z$ axis, ${\vec B} = (0,0,B)$.
Throughout the paper  rydberg (Ry) is used as the energy unit.
For the magnetic field we use either atomic units or Gauss (G)
with the conversion factor $B_0=1$\,a.u. $=2.35\times 10^9$\,G.
For the other quantities standard atomic units are used. The
distances between infinitely-heavy charged centers are denoted by
$R$ letters, whereas the distances between centers and electron are
denoted by $r$ letters. The distance between the electron
position and the $z$ axis is denoted by $\rho$. In particular, the
potential corresponding to the hydrogen atom is given by
\begin{equation}
\label{eq:1}
 V\ =\ -\frac{2}{r}  + \frac{B^2 \rho^2}{4} \ ,
\end{equation}
where $\rho = \sqrt{x^2+y^2}$ and $r$ is the distance from the
electron to the charged center. The potential
\begin{equation}
\label{H2+}
 V\ =\ \frac{2 Z_1 Z_2}{R} -\frac{2 Z_1}{r_1} -\frac{2 Z_2}{r_2} +
 \frac{B^2 \rho^2}{4} \ ,
\end{equation}
describes the ions ${\rm H}_2^{+}$ (the system $({\rm ppe}), Z_1=Z_2=1$),\
$({\rm HeH})^{++}$\ (the system $(\al {\rm p e})$, $Z_1=1,Z_2=2$), ${\rm He}_2^{(3+)}$\
(the system $(\al \al {\rm e}), Z_1=Z_2=2$), where $r_1 (r_2)$ is the
distance from the electron to the charged center $1$ ($2$) and $R$ is
the distance between the charged centers. In turn, the system
${\rm H}_3^{++}$ is described by the potential
\begin{equation}
\label{H3++}
 V\ =\ \frac{2}{R_+} + \frac{2}{R_ -}
+ \frac{2}{R_+ + R_-} -\frac{2}{r_1} -\frac{2}{r_2} -\frac{2}{r_3}
  +  \frac{B^2 \rho^2}{4} \ ,
\end{equation}
where $r_i$ is the distance from the electron to the charged center
$i$ and $R_{\pm}$ are the distances from the central charge, placed in
the origin, and the side charged centers. The  equilibrium distance,
which corresponds to the minimum of the total energy, is defined by
the distance between the most-distant charged centers, which is
$L_{\rm eq}=R_{\rm eq}$ for the two-center case of ${\rm H}_2^+,
 ({\rm HeH})^{++}$, ${\rm He}_2^{3+}$ and
$L_{\rm eq}={R_+}_{\rm eq} + {R_-}_{\rm eq}$ for three-center case of ${\rm H}_3^{++}$. Hereafter,
magnetic field is defined in dimensionless units (a.u.) as $B/B_0$,
where $B_0 = 2.35 \times 10^9$\,G, which we continue to denote as $B$.

\section{The ${\rm H}$-atom}

Let us take hydrogen atom - the simplest one-electron system -
placed in a constant uniform magnetic field $B$ directed along the
$z$-axis. Due to the Lorentz force, the spherical symmetric
electron cloud (in the absence of a magnetic field) is deformed to a
cigar-like form. The size of the electron cloud $r_{\rm t}$ in
transversal direction to $z$-axis shrinks drastically $\sim
B^{-1/2}$ at large magnetic fields, being close the value of the
Larmor radius. As to the longitudinal size $r_{\rm l}$ it also
contracts at large magnetic fields but at a much more moderate rate,
$\sim {(\log B)}^{-1}$ (see, e.g., \cite{Ruderman:1971,LL-QM:1977,Hasegawa:1961}).
An interplay of these
two types of  behavior explains the cigar-type form of the
electron cloud. At very large magnetic fields, the cigar-type form
evolves to a needle-like form known as the {\it Ruderman needle}.
In Fig.~1a the form of the electron cloud is illustrated for
$B=10^{12}$\,G \footnote{Calculations were made using the trial
function (7) from \cite{Potekhin:2001}}. In particular, the
longitudinal size of the electron cloud shrinks in comparison
with the zero-magnetic-field case  about four times. Therefore, the
apparent classical (electrostatic) appearance of the magnetic
field influence is characterized by a change of the form of the
electron cloud, which can be roughly approximated by the ratio
of two classical parameters $r_{\rm t}, r_{\rm l}$. In fact, it is the major
assumption of the present approximation scheme. We also assume
that these parameters $r_{\rm t}, r_{\rm l}$ are defined by the expectation
values,
\begin{equation}
\label{pars}
    r_{\rm t} \equiv <\rho>\ ,\ r_{\rm l} \equiv 2<|z|>  \ .
\end{equation}
If a definition of the transversal size $r_{\rm t}$ looks natural from
the physical point of view and rather unambiguous,  definition
(\ref{pars}) of the longitudinal size is not so obvious. It can be
chosen as $\sim \sqrt{<z^2>}$, or as a linear combination of
$<|z|>$ and $\sqrt{<z^2>}$. So far it is not so clear what would
be physical arguments which allow to specify a definition.
Eventually, it turns out it is not very important what quantity is
used to define $r_{\rm l}$. The results of the fit remain very
similar although there can be some difference in the parameters.

\begin{figure}
\begin{center}
  \includegraphics*[width=3.in,angle=0]{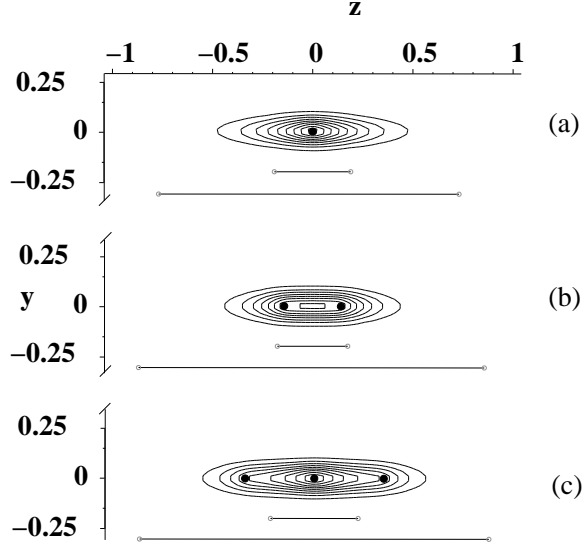}
    \caption{The contours of the electron clouds on the plane $x=0$ for the
      ${\rm H}$ atom (a), ${\rm H}_2^+$ ion (b) and ${\rm H}_3^{++}$ (c) placed in a
      magnetic field $B=10^{12}$\,G directed along the $z$-axis, $(x,y,z)$ are in a.u. For
      the ${\rm H}_2^+, {\rm H}_3^{++}$-ions the protons are situated on the
      $z$-line. The values of $2<|z|>$ are shown by bars: the short
      bars correspond to $B=10^{12}$\,G , they are compared with the
      long bars which correspond to $B=0$ for ${\rm H}, {\rm H}_2^+$ and to
      $B=10^{10}$\,G (near the threshold of existence) for ${\rm H}_3^{++}$.
      It illustrates a shrinking of electronic longitudinal size with
      magnetic field growth.}
  \label{fig:6.1}
\end{center}
\end{figure}

The binding energy $E_{\rm b}$ is by definition the difference between
the energy of free electron in magnetic field (the Larmor energy)
$B$ and the total energy of the atom, $E_{\rm b}=B-E_{\rm T}$. It is known
that $E_{\rm b}$ in the  weak-field regime is represented by the Taylor
expansion in powers of $B^2$, while for large $B$ it behaves $\sim
(\log^2 B)$ (see, for example, \cite{LL-QM:1977} and discussion in
\cite{Karnakov:2003}). Following the above assumption the binding
energy depends on the ratio ${\X}=r_{\rm t}/r_{\rm l}$,
\begin{equation}
\label{eb}
    E_{\rm b}\ =\ E_{\rm b} ({\X})\ .
\end{equation}

It is quite natural to approximate the transverse size $r_{\rm t} \equiv
<\rho>$ as follows
\begin{equation}
\label{rt}
    r_{\rm t}= \frac{r^0_{\rm t}}{(1 + \al_{\rm t}^2 B^2)^{1/4}}
    \bigg(\frac{1+a_{\rm t} B^2}{1+b_{\rm t} B^2}\bigg)\ ,
\end{equation}
where $r^0_{\rm t}, \al_{\rm t}, a_{\rm t}, b_{\rm t}$ are parameters,
which are found by fitting the calculated expectation values for
$<\rho>$. The formula (\ref{rt}) is written in such a way as to
reproduce a functionally-correct perturbative expansion of $<\rho>$ at
$B=0$ (in powers $B^2$) and $r_{\rm t}^0 = \sqrt{2}a_{\rm B}$, where
$a_{\rm B}=1$\,a.u.  is the Bohr radius. At large $B$ the right-hand
side of Eq.(\ref{rt}) behaves as $\sim B^{-1/2}$ simulating the Larmor
radius behavior.

\begin{figure}
\begin{center}
   \includegraphics*[width=2.25in,angle=-90]{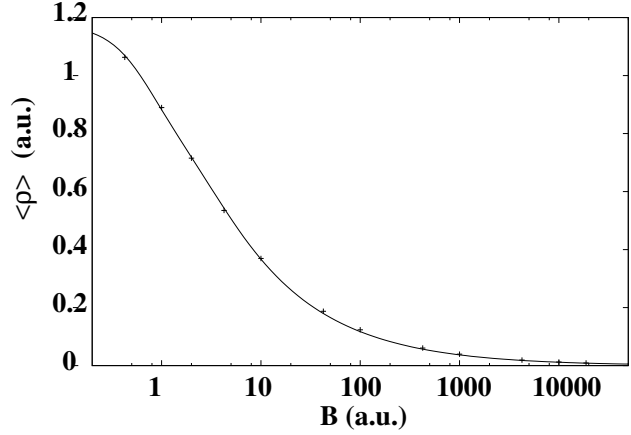}
    \caption{${\rm H}$ atom: the fit of transverse size of the
     electron cloud $<\rho>$ versus $B$ using Eq.(\ref{rt}).
     Calculated values are indicated by crosses. }
   \label{fig:6.2}
\end{center}
\end{figure}

\noindent In Fig.~\ref{fig:6.2} one can see that Eq.(\ref{rt})
fits data on $<\rho>$ calculated using the formalism developed in
\cite{Potekhin:2001} with accuracy better than one percent at
$B\gtrsim 10^9$\,G. The parameters of the fit are given in
Table~\ref{table:6.1}. The parameter $\ r^0_{\rm t}$ is also found from
the fit, it deviates from $\sqrt{2}$ (see above) by $\simeq 8\%$. It
reflects the fact that the accuracy provided by  formula
(\ref{rt}) diminishes as the magnetic field decreases (see the
discussion below).

\begin{table*}[htbp]
  \centering
    \caption{
  The parameters of the fit (\ref{rt}) of the transversal size
  of the electron clouds of the ${\rm H}$ atom and
  ${\rm H}_2^+, {\rm H}_3^{++}, ({\rm HeH})^{++}, {\rm He}_2^{3+}$ ions (in a.u.).
  }
\begin{tabular}{|c||c|c|c|c|}\hline
 System     & $\ r^0_{\rm t}$ & $\ \al_{\rm t}$  &\ $a_{\rm t}$    &\ $b_{\rm t}$\ \\
\hline \hline
 \ ${\rm H}$ atom\ &\ 1.17533\ &\ 0.44904\ &\ 1.20981\ &\ 1.81098\ \\
\hline
 \ ${\rm H}_2^+$\ &\ 0.954427\ &\ 0.23615\ &\ 0.376237\ &\ 0.62194\ \\
\hline
 \ ${\rm H}_3^{++}$\ &\ 0.645875\ &\ 0.048196 \ &\ 0.00970609\ &\ 0.0230488\ \\
\hline \hline
 \ $({\rm HeH})^{++}$\ &\ 0.174416\ &\ 0.019657 \ &\ 0.00000030\ &\ 0.0000003\ \\
\hline \hline
 \ ${\rm He}_2^{3+}$\ &\ 0.200825 \ &\ 0.026449 \ &\ 0.00000150\ &\ 0.00000148\ \\
\hline
\end{tabular}
  \label{table:6.1}
\end{table*}

At first sight, it is a much more complicated task to describe the
longitudinal size, $\nobreak{r_{\rm l}\equiv 2<|z|>}$. The
approximation we propose to use is
\begin{equation}
\label{rl}
  r_{\rm l}= \frac{r^0_{\rm l}}{1 + \al_{\rm l} \log(1+\beta_{\rm l}^2 B^2+\gamma_{\rm l}^2 B^4)}
  \bigg(\frac{1+a_{\rm l} B^2}{1+b_{\rm l} B^2}\bigg)\ ,
\end{equation}
where $r^0_{\rm l}, \al_{\rm l}, \beta_{\rm l}, \gamma_{\rm l}, a_{\rm
  l}, b_{\rm l}$ are parameters, which are found by fitting the
calculated expectation values for $2<|z|>$. Formula (\ref{rl}) has the
perturbative expansion in powers $B^2$, which agrees with perturbation
theory results and $r_{\rm l}^0 = 3/2\, a_B$, where $a_{\rm
  B}=1\,$a.u. is the Bohr radius. At large $B$, the right-hand side
(\ref{rl}) behaves as $\sim {(\log B)}^{-1}$ as should be in
accordance with the qualitative arguments.

\begin{figure}
\begin{center}
   \includegraphics*[width=2.25in,angle=-90]{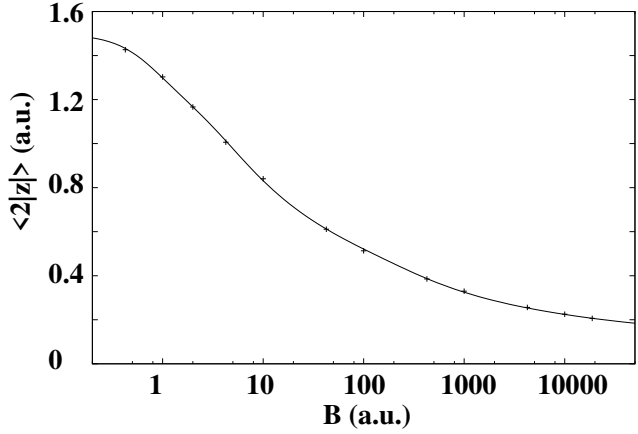}
    \caption{${\rm H}$ atom: the fit of longitudinal size of the
    electronic cloud $2<|z|>$ versus $B$ using Eq.(\ref{rl}).
    Calculated values are indicated by crosses.}
   \label{fig:6.3}
\end{center}
\end{figure}

In Fig.~\ref{fig:6.3} one can see that (\ref{rl}) fits data on
$2<|z|>$ obtained in the formalism developed in \cite{Potekhin:2001}
with accuracy better than 1\% at $B\gtrsim 10^9$\,G. The parameters of
the fit are given in Table~\ref{table:6.2}. The parameter $r^0_{\rm
  l}$ is also found from fit. Surprisingly, it deviates from $3/2$
(see above) insignificantly, by $\lesssim 1\,\%$, in contrast to
what happened for the parameter $r^0_{\rm t}$.

\begin{figure}
\begin{center}
   \includegraphics*[width=2.25in,angle=-90]{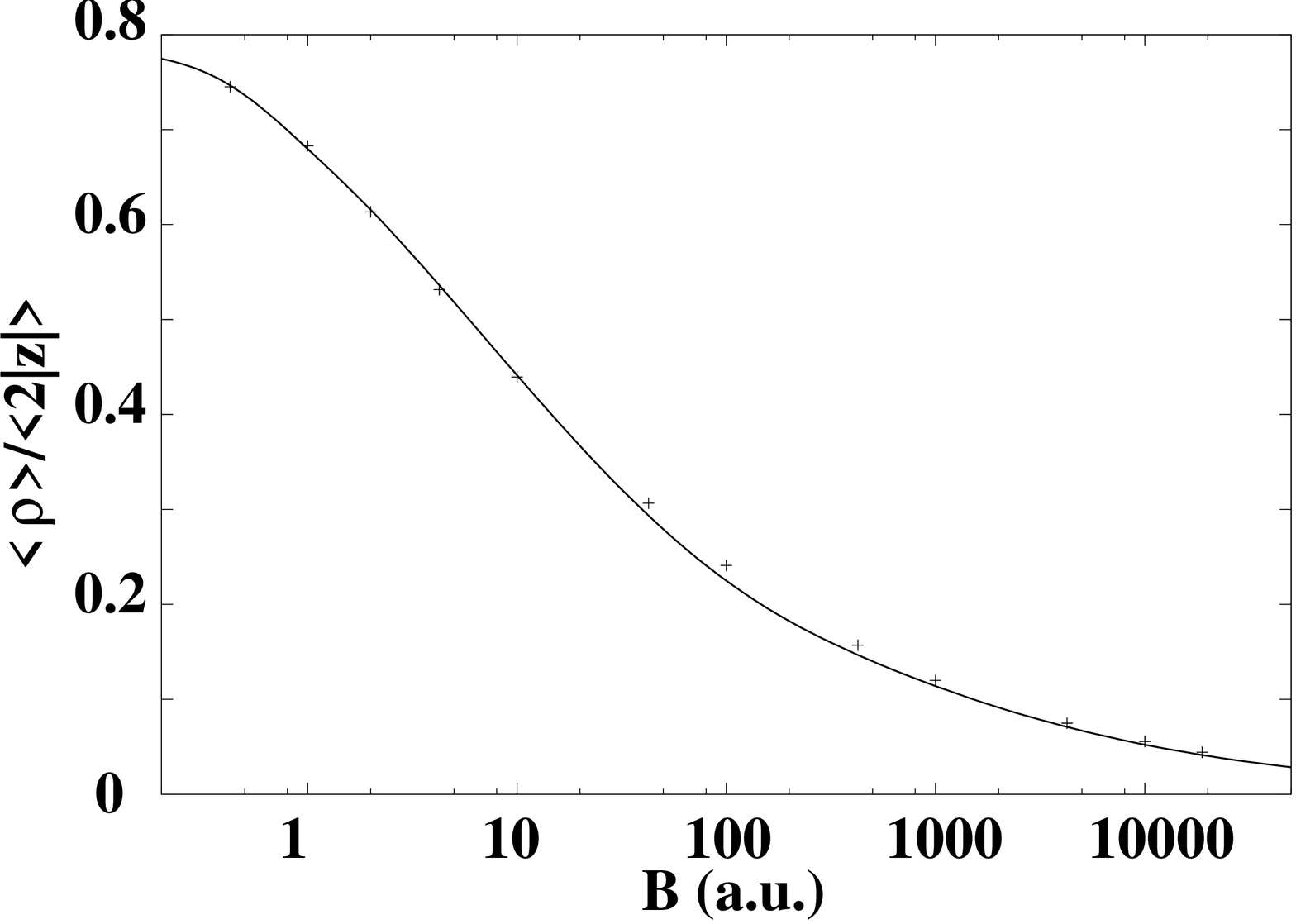}
    \caption{${\rm H}$ atom: ratio  \(\X = \frac{<\rho>}{2<\vert z\vert>}\).
    Calculated values are indicated by crosses.}
   \label{fig:6.4}
\end{center}
\end{figure}

In Fig.~\ref{fig:6.4} a comparison of the ratio ${\X} = r_{\rm
  t}/r_{\rm l}$ (see Eqs. (\ref{rt})-(\ref{rl})) with parameters taken
from Table~\ref{table:6.1} with results of calculations is presented.
One can clearly see that both data and fitted curves of
Figs.~\ref{fig:6.3} and \ref{fig:6.4} demonstrate a certain
irregularity in the range $(5-50) \times 10^{10}$\,G. It is a
transition region from the Coulomb regime, where the Coulombic forces
dominate over magnetic forces to the Landau regime where in the $(x,y)$
plane Coulombic forces become subdominant.

\begingroup
\begin{table*}[htbp]
  \centering
    \caption{
  The parameters of the fit of the longitudinal size of the
  electron clouds of ${\rm H}$ atom, ${\rm H}_2^+, {\rm H}_3^{++}$ molecular
  ions as well as $({\rm HeH})^{++}, {\rm He}_2^{3+}$
  using by (\ref{rl}) (in a.u.).
 }
  \begin{tabular}{|c||c|c|c|c|c|c|}\hline
System    & $r^0_{\rm l}$& $\al_{\rm l}$ & $\beta_{\rm l}$ & $\gamma_{\rm l}$ & $a_{\rm l}$ &
$b_{\rm l}$
 \\ \hline \hline
\ ${\rm H}$ atom\ &\ 1.49719\ &\ 0.179332\ &\ 0.320252\ &\ 0.001164\ &\
1.07512\ &\ 1.35162\
 \\ \hline
\ ${\rm H}_2^+$\ &\ 1.72041\ &\ 0.254255\ &\ 0.141004\ &\ 0.0004436\ &\
0.340131\ & 0.497807\
 \\ \hline
\ ${\rm H}_3^{++}$\ &\ 1.94408\ &\ 0.279956\ &\ 0.0191558\   &\
0.000008\ &\ 0.0066712\ &\ 0.0136894\
 \\ \hline \hline
\ $({\rm HeH})^{++}$\ &\ 1.72219 \ &\ 1.155934\ &\ 0.405312\ &\
0.000311\ &\ 0.131457\ &\ 0.0600481\
 \\ \hline \hline
\ ${\rm He}_2^{3+}$\ &\ 0.65727\ &\ 0.228616 \ &\ 0.0011334\ &\ 0.\ &\
0.0000116\ &\ 0.0000228\
 \\ \hline
  \end{tabular}
 \label{table:6.2}
\end{table*}
\endgroup

Following the assumption (\ref{eb}) let us approximate the binding
energy
\begin{equation}
\label{BE}
    E_b = A {\X}_{\rm l}^2 + B {\X}_{\rm l} + C \ ,\ {\X}_{\rm l}=\log {\X} \ ,
\end{equation}
where $A,B,C$ are parameters, which are found by making fit of the
results of calculations of the binding energy. It is worth
emphasizing that the parameters of ${\X}({\X}_{\rm l})$ are already
fixed by the fits (\ref{rt}), (\ref{rl}) of $r_{\rm t}, r_{\rm l}$,
respectively. The formula (\ref{BE}) agrees with the perturbative
expansion in powers $B^2$ (at small $B$) and gives a correct
asymptotic expansion at large $B$.

\begin{figure}
\begin{center}
  \includegraphics*[width=2.25in,angle=-90]{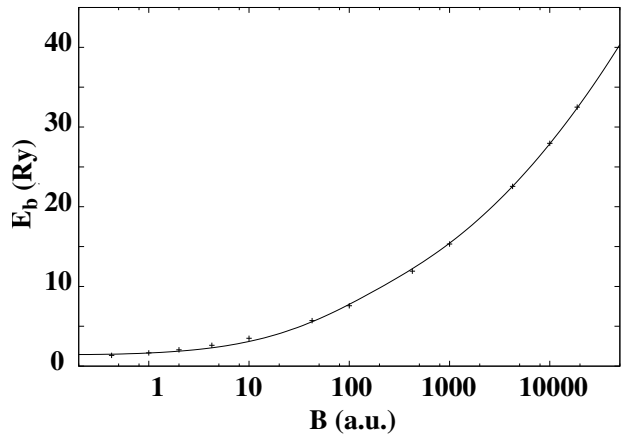}
  \caption{${\rm H}$ atom: the fit of the binding energy using (\ref{BE}).
  Calculated values are indicated by crosses.}
  \label{fig:6.5}
\end{center}
\end{figure}
In Fig.~\ref{fig:6.5} it is shown the fit using the formula (\ref{BE})
of the best known results for the binding energies from
\cite{Kravchenko:1996} combined with those from \cite{Potekhin:2001}.
The parameters $A,B,C$ are given in Table~\ref{table:6.3}. In the
whole range of explored magnetic fields $10^9 - 4.414 \times
10^{13}$\,G, formula (\ref{BE}) approximates the binding energies with
a relative accuracy, which does not exceed few percent and becomes
more accurate with growing magnetic field. It is worth mentioning that
when the parameter $A=4$ in the approximation (\ref{BE}) its
asymptotic coincides with the exact asymptotic (see, e.g.,
\cite{LL-QM:1977}, \S 112),
\begin{equation}
\label{As-Binding}
 E_{\rm b} \approx \log^2 \frac{B}{B_0}\ ,\ B \rar \infty\ ,
\end{equation}
where $E_{\rm b}$ is in Ry. In fact, the deviation $|A/4-1|$ gives a
feeling about the quality of our approximation. Clearly, this estimate
is very rough ca. 20\,\% (see Table III), while a real accuracy
of approximating the binding energy is a few percent.

\begin{table}[htbp]
  \centering
    \caption{
  The parameters of the fit of the binding energy of the
  ${\rm H}$ atom, ${\rm H}_2^+, {\rm H}_3^{++}$ molecular ions as well as
  $({\rm HeH})^{++}, {\rm He}_2^{3+}$ using (\ref{BE}) (all in Ry).
  }
  \begin{tabular}{|c||c|c|c|}\hline
System    &\ $A$ \ &\ $B$\ &\ $C$\
   \\ \hline \hline
\ ${\rm H}$-atom\ &\ 3.22532 &\ 0.53945 \ &\ 1.37932 \
  \\ \hline
\ ${\rm H}_2^+$\ &\ 8.23442\ &\ 6.8246\ &\ 2.99945\
  \\ \hline
\ ${\rm H}_3^{++}$\ &\ 12.8455\ &\ 20.4849\ &\ 3.95821\
  \\ \hline
\ $({\rm HeH})^{++}$\ &\ 15.7401\ &\ 6.1134\ &\ -5.3756\
  \\ \hline
\ ${\rm He}_2^{3+}$\ &\ 26.2926 \ &\ 32.9181\ &\ -0.28129\
  \\ \hline

  \end{tabular}
  \label{table:6.3}
\end{table}

\begin{figure}
\begin{center}
   \includegraphics*[width=2.25in,angle=-90]{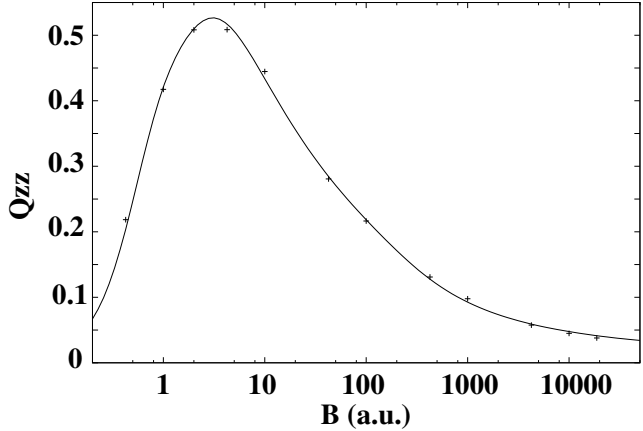}
    \caption{${\rm H}$ atom: the fit of the quadrupole moment using (\ref{QF}).
    Calculated values (see \cite{Potekhin:2001}) are indicated by
    crosses.}
   \label{fig:6.6}
\end{center}
\end{figure}

One of the important characteristics of the magnetic field
influence on the ${\rm H}$-atom is the appearance of the quadrupole
moment
\begin{equation}
\label{q-def}
   Q \equiv - Q_{zz} = 2 <z^2> - <\rho^2>\ .
\end{equation}
Recently, the first quantitative study of the quadrupole moment was
carried out \cite{Potekhin:2001}. The formula (\ref{q-def}) suggests
immediately the following approximation
\begin{equation}
\label{QF}
    Q = 2 r_{\rm l}^2 (A_q - a_q {\X}_{\rm l}) - r_{\rm t}^2 (B_q + b_q {\X}_{\rm l}) \ ,\
    {\X}_{\rm l}= \log {\X} \quad ,
\end{equation}
where $A_q=0.325447, a_q=0.049432, B_q=1.32012, b_q=0.955362$ are
dimensionless parameters, which are found by fitting the quadrupole
moment. The parameters of ${\X}_{\rm l}$ are already fixed in the fit
of parameters $r_{\rm t}, r_{\rm l}$ using (\ref{rt}) and (\ref{rl}),
respectively. Formula (\ref{QF}) describes correctly the expansion at
small and large $B$ (see
\cite{Ruderman:1971,Turbiner:1987,Potekhin:2001}). It fits the results
of calculations in \cite{Potekhin:2001} with an accuracy of few
percents (see Fig.~\ref{fig:6.6}).

We made an analysis of the expectation values $<|z|^n>$ at
$n=2,3,4,5$. It turns out that the calculated expectation values
admit a very accurate polynomial approximation in terms of a single
expectation value $<|z|>$,
\begin{equation}
\label{expect}
    <|z|^n>\ = \ P_n (<|z|>)  \ ,
\end{equation}
where $P_n$ is a $n$-th degree polynomial. It seems natural to
assume that (\ref{expect}) holds for any $n$, hence any expectation
value is defined by $<|z|>$. This leads to a
striking hypothesis that the ground state eigenfunction integrated
over $\rho$ can be viewed as a one-parametric probability distribution (!).

\section{The ${\rm H}_2^+$ molecular ion}

In this Section we consider the molecular ion ${\rm H}_2^+$ in parallel
configuration, when the protons are situated along the magnetic line.
The form of the electron cloud is illustrated in Fig.~1b for the
magnetic field $B=10^{12}$\,G. The transversal size of the electron
cloud $r_{\rm t}$ shrinks drastically, $\sim B^{-1/2}$, at large
magnetic fields, being close to the value of the Larmor radius
similarly to what happens for the hydrogen atom. As to the
longitudinal size $r_{\rm l}$ it also shrinks but at a much slower
rate $\sim {(\log B)}^{-1}$. In particular, the longitudinal size of
the electron cloud shrinks in comparison with vanishing magnetic field
about five times (see Fig.~1b).

\begin{figure}
\begin{center}
   \includegraphics*[width=2.25in,angle=-90]{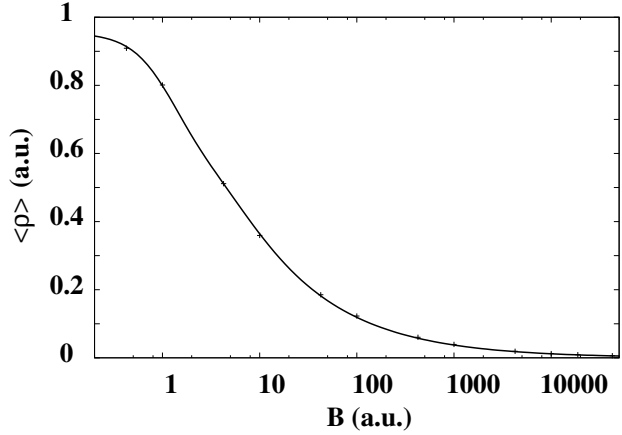}
    \caption{${\rm H}_2^+$ ion: a fit of the transverse size of the
     electron cloud $<\rho>$ using Eq.(\ref{rt}).
     Calculated values are indicated by crosses.}
   \label{fig:6.7}
\end{center}
\end{figure}

\begin{figure}
\begin{center}
   \includegraphics*[width=2.25in,angle=-90]{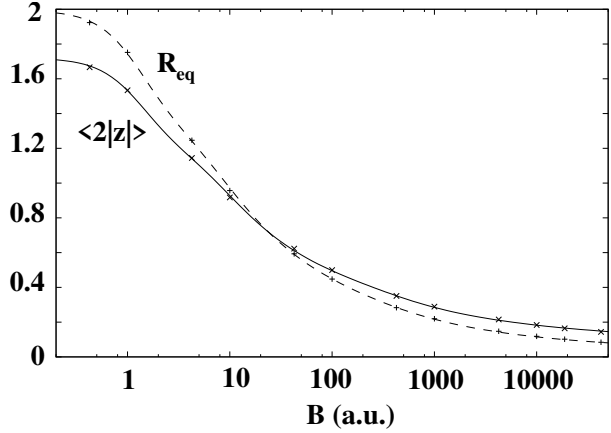}
    \caption{${\rm H}_2^+$ ion: a fit of the longitudinal size of the
      electron cloud $2<|z|>$ using Eq.(\ref{rl}) (solid curve) and of
      the equilibrium distance $R_{\rm eq}$ (dashed curve) using
      Eq.(\ref{Req}). Calculated values are indicated by crosses.}
   \label{fig:6.8}
\end{center}
\end{figure}

\begin{figure}
\begin{center}
   \includegraphics*[width=2.25in,angle=-90]{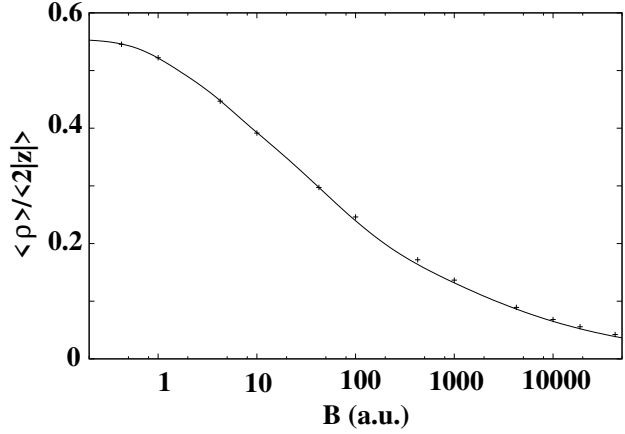}
    \caption{${\rm H}_2^+$ ion: the ratio  ${\X} = \frac{<\rho>}{2<|z|>}$.
    Calculated values are indicated by crosses.}
   \label{fig:6.9}
\end{center}
\end{figure}

\begin{figure}
\begin{center}
   \includegraphics*[width=2.25in,angle=-90]{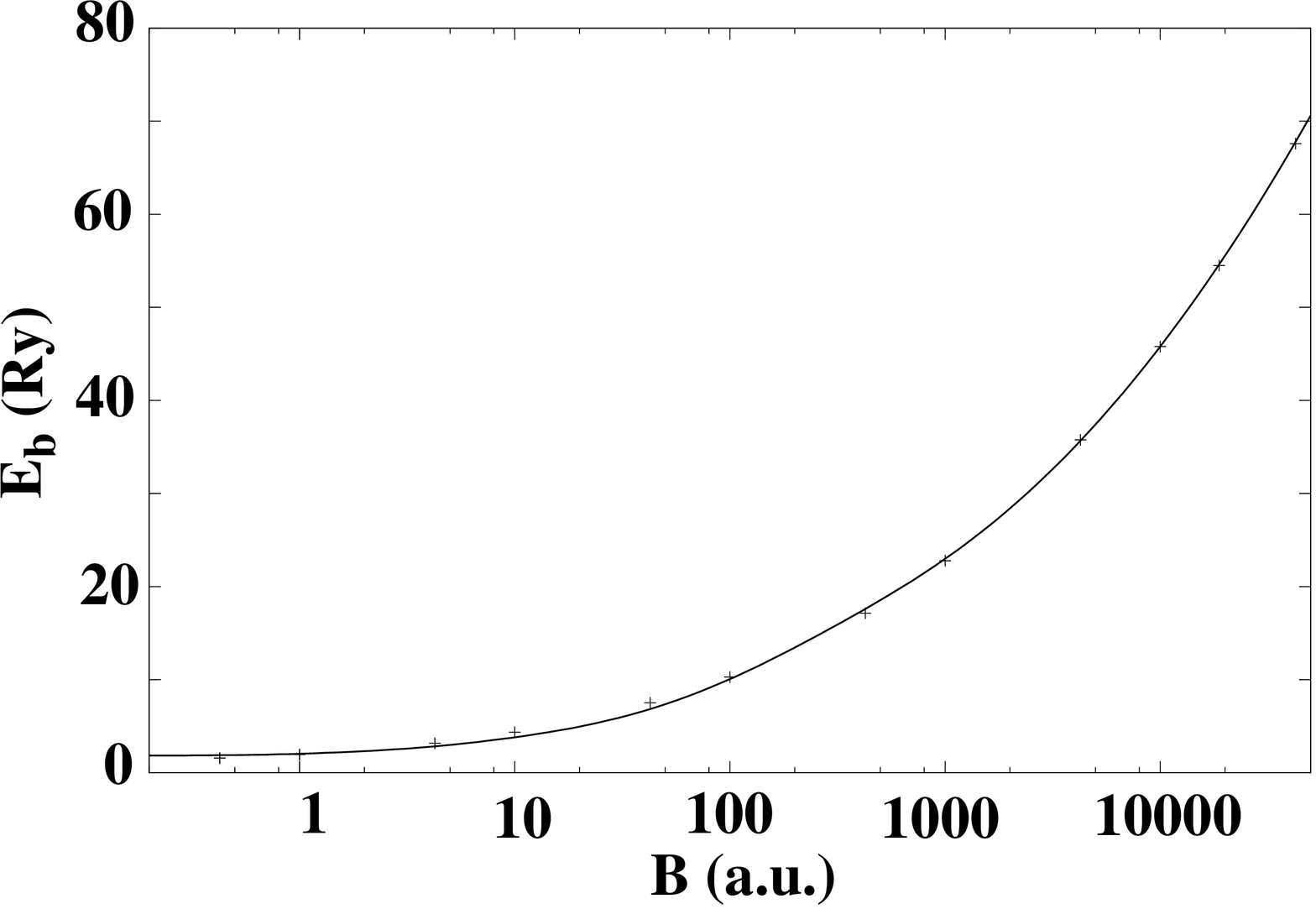}
    \caption{${\rm H}_2^+$ ion: the fit of the binding energy using
    (\ref{BE}). Calculated values are indicated by crosses.}
   \label{fig:6.10}
\end{center}
\end{figure}

Following the same arguments which were used earlier for the ${\rm H}$ atom,
we again assume that the dynamic characteristics of the ${\rm H}_2^+$ in a
magnetic field depend on the expectation values of transversal
$(r_{\rm t})$ and longitudinal $(r_{\rm l})$ sizes. The dependence of
them on magnetic field is approximated by similar formulas
(\ref{rt})-(\ref{rl}). The binding energy at equilibrium distance
between protons depends on the ratio ${\X}=r_{\rm t}/r_{\rm l}$.
Eventually, the binding energy is written in the same form (\ref{BE})
with the same expressions (\ref{rt}) and (\ref{rl}) as is done for ${\rm H}$
atom but with different parameters. For the fit we use the results of
recent calculations of the binding energy which were carried out in
\cite{Turbiner:2003,Turbiner:2004}. These parameters of the fit are
presented in Tables~\ref{table:6.1}-\ref{table:6.3} and the fit is
illustrated by Figs.~\ref{fig:6.7}-~\ref{fig:6.10}.

In order to approximate the equilibrium distance $R_{\rm eq}$ we
assume that $R_{\rm eq}$ is proportional to the longitudinal distance
$r_{\rm l}$ with a small correction in ${\X}_{\rm l}$
\begin{equation}
\label{Req}
  R_{\rm eq}= r_{\rm l} (c_0 + c_1 {\X}_{\rm l} + c_2 {\X}_{\rm l}^2)=
   \frac{r^0_{\rm l}}{1 + \al_{\rm l} \log(1+\beta_{\rm l} B^2)}
    \bigg(\frac{1+a_{\rm l} B^2}{1+b_{\rm l} B^2}\bigg)
(c_0 + c_1 {\X}_{\rm l} +
    c_2 {\X}_{\rm l}^2)\ ,
\end{equation}
where the parameters $c_0, c_1, c_2$ are found from the fit of the
results of calculations of the equilibrium distance which were carried
out in \cite{Turbiner:2003,Turbiner:2004}.  The parameters of the fit
are given in Table~\ref{table:6.4}. The fit is illustrated in
Fig.~\ref{fig:6.8}. It is worth mentioning that the parameters $c_i,\
i=0,1,2$ decrease very fast with $i$ (see Table~\ref{table:6.4}). This
can be considered as an indication of adequateness of the
approximation formula (\ref{Req}).

\begin{table}[htbp]
  \centering
    \caption{  The dimensionless parameters of the fit (\ref{Req})
    of the equilibrium distance (in a.u.) of ${\rm H}_2^+, {\rm H}_3^{++}$ and
    $({\rm HeH})^{++}, {\rm He}_2^{3+}$ions .}
  \begin{tabular}{|c||c|c|c|}\hline
System    & $c_0$      & $c_1$    & $c_2$
 \\ \hline \hline
${\rm H}_2^+$   &\ 1.37384\ &\ 0.389879\  &\ 0.0430844
 \\ \hline
${\rm H}_3^{++}$&\ 4.48200\ &\ 2.25814\   &\ 0.380948
 \\ \hline \hline
$({\rm HeH})^{++}$&\ 4.15754\ &\ 2.31113 \ &\ 0.409048
 \\ \hline \hline
${\rm He}_2^{3+}$ &\ 1.83774 \ & 0.51165 \ &\ 0.0626179
 \\ \hline
  \end{tabular}
  \label{table:6.4}
\end{table}

Similarly to what happened for ${\rm H}$ atom, the plot of the ratio ${\X} =
r_{\rm t}/r_{\rm l}$ (Fig.~\ref{fig:6.9}) reveals a certain
irregularity in behavior of the calculation results as well as the fit
in the range $(5-50) \times 10^{10}$\,G. We assign these
irregularities to a transition from the Coulomb to the Landau regime.
An overall quality of the fit for the domain $10^9 - 4\times
10^{13}$\,G is very high, about 1-2\,\% except for the above-mentioned
region where the accuracy drops  to 5-10\,\%.

Similar to what was done for the ${\rm H}$-atom we carried out a calculation
of expectation values $<|z|^n>~$, $n=2,3,4,5$. It turns out that these
expectation values admit very accurate polynomial approximation in
terms of the expectation value $<|z|>$ (see (\ref{expect})). It seems
natural to assume that (\ref{expect}) holds for any $n$. This leads to
the hypothesis that the ground state eigenfunction integrated over
$\rho$ defines a one-parametric distribution similar to what appears
for the ${\rm H}$ atom (see previous Section).

\section{The ${\rm H}_3^{++}$ molecular ion}

Now we consider the exotic system ${\rm H}_3^{++}$ theoretically predicted
in \cite{Turbiner:1999}, which is made out of three protons situated
along the magnetic line and one electron (parallel configuration).
This system appears as a quasi-stationary state at $B \gtrsim
10^{10}$\,G \cite{Turbiner:2004}. The form of the electron cloud for
$B=10^{12}$\,G is shown in Fig.~1c. It is clearly seen that the
transversal size of the electron cloud $r_{\rm t}$ shrinks
drastically, $\sim B^{-1/2}$ at large magnetic fields similar to what
happens for the hydrogen atom and the ${\rm H}_2^+$ molecular ion which is
of the order of the Larmor radius. As to the longitudinal size $r_{\rm
  l}$ it also contracts but in much slower rate, $\sim {(\log
  B)}^{-1}$, at large magnetic fields.

\begin{figure}
\begin{center}
   \includegraphics*[width=2.25in,angle=-90]{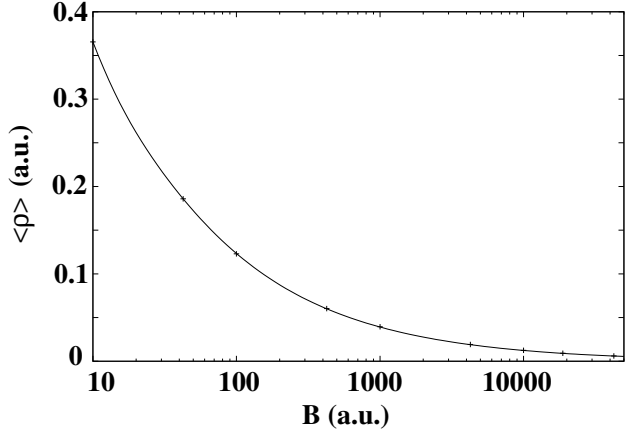}
    \caption{${\rm H}_3^{++}$ ion: the fit of transverse size of the
    electron cloud $<\rho>$ using Eq.(\ref{rt}).
    Calculated values are indicated by crosses.}
   \label{fig:6.11}
\end{center}
\end{figure}

\begin{figure}
\begin{center}
   \includegraphics*[width=2.25in,angle=-90]{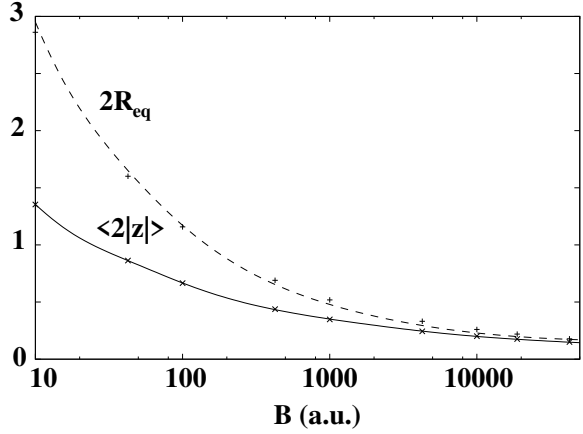}
    \caption{${\rm H}_3^{++}$ ion: the fit of longitudinal size of the
      electron cloud $2<|z|>$ using Eq.(\ref{rl}) (solid line) and
      of the equilibrium distance $L_{\rm eq}=2 R_{\rm eq}$ using
      Eq.(\ref{Req}) (dashed line).  Calculated values are indicated
      by crosses.}
   \label{fig:6.12}
\end{center}
\end{figure}

\begin{figure}
\begin{center}
   \includegraphics*[width=2.5in,angle=-90]{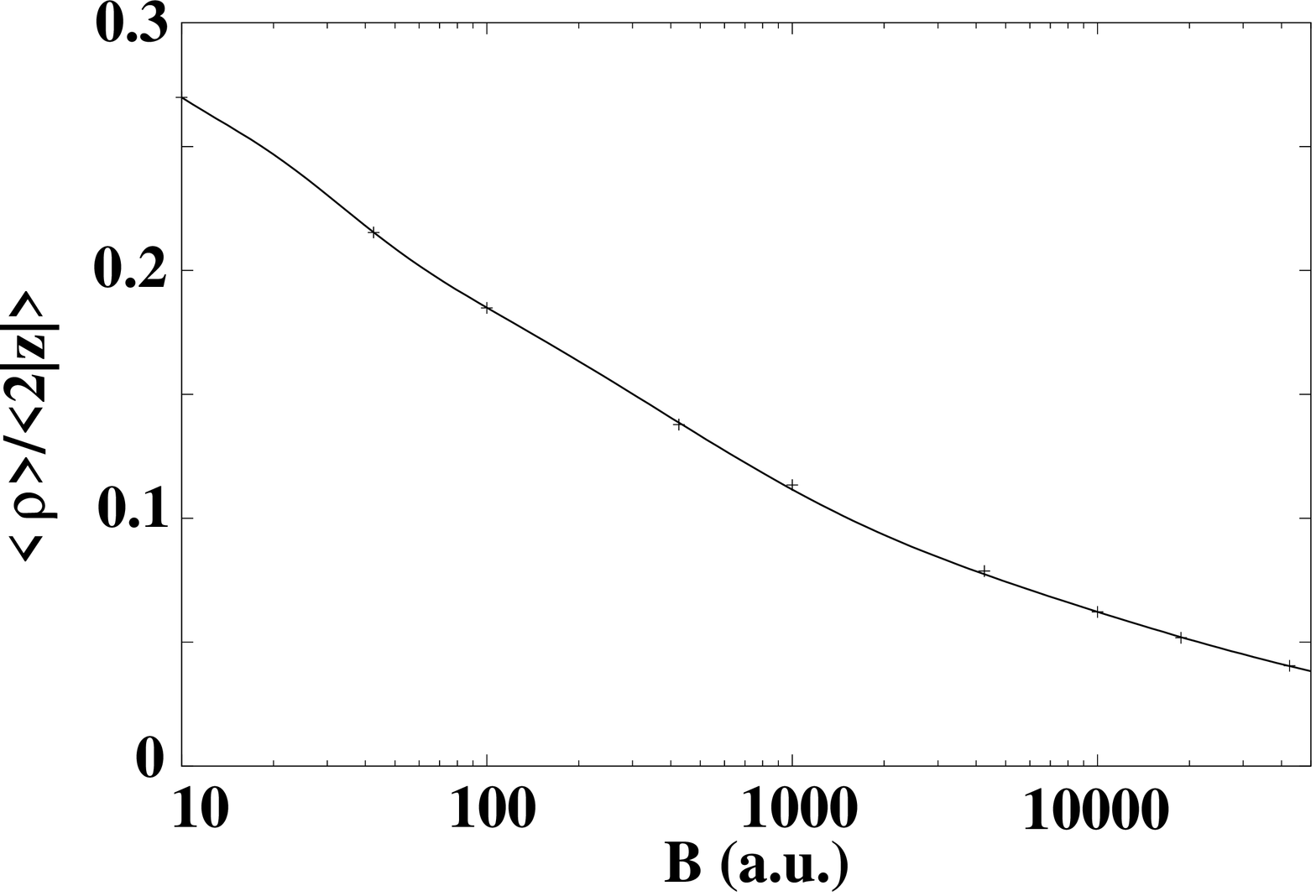}
    \caption{${\rm H}_3^{++}$ ion: the ratio  ${\X}=\frac{<\rho>}{2<|z|>}$.
    Calculated values are indicated by crosses.}
   \label{fig:6.13}
\end{center}
\end{figure}

\begin{figure}
\begin{center}
   \includegraphics*[width=2.5in,angle=-90]{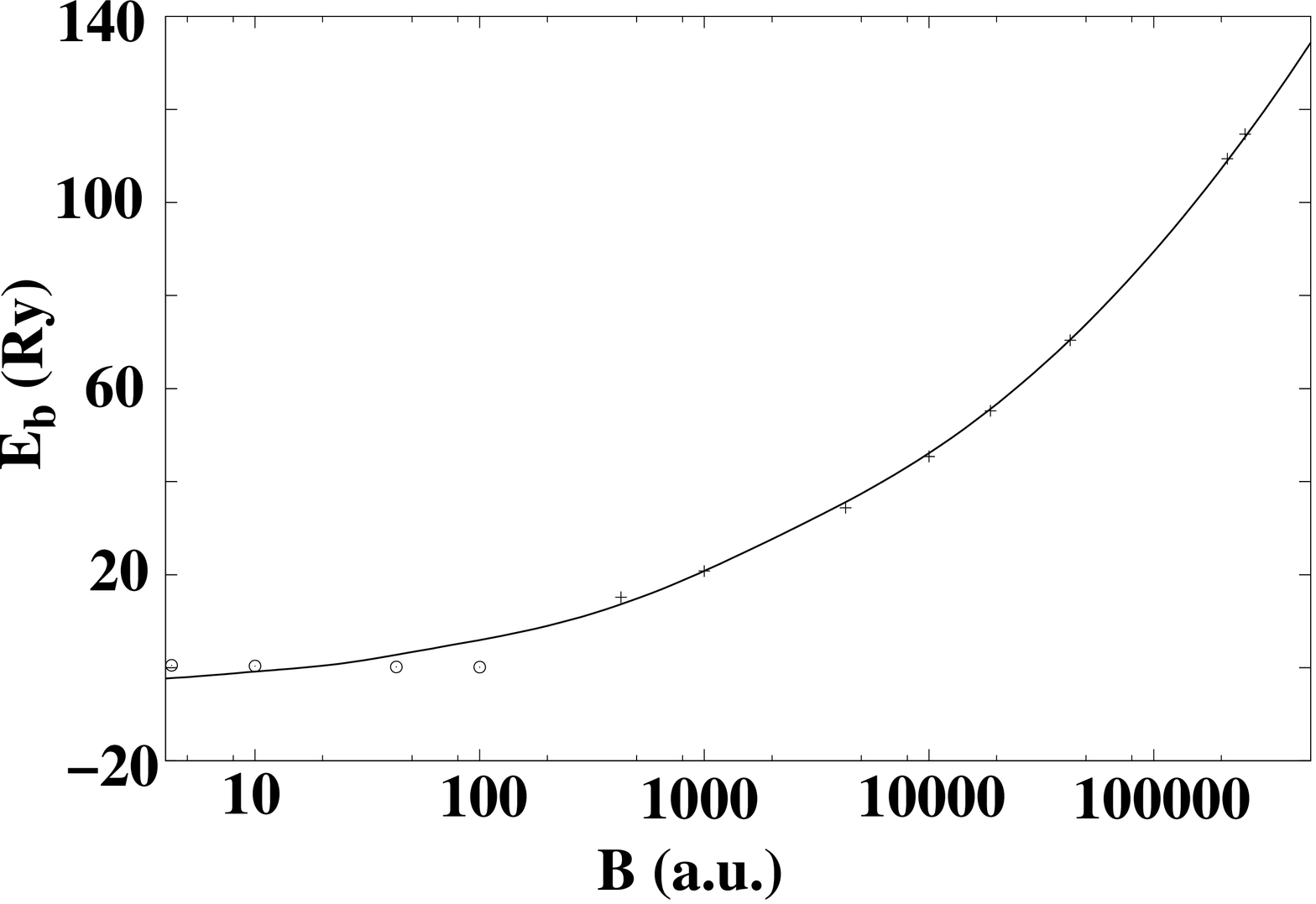}
    \caption{${\rm H}_3^{++}$ ion: the fit of the binding energy using
      (\ref{BE}). Only calculated values which are indicated by
      crosses are used for fitting, while calculated values shown by
      circles are not taken into account (see text).  }
   \label{fig:6.14}
\end{center}
\end{figure}

We follow the same idea of approximation as for ${\rm H}$ and ${\rm H}_2^+$
assuming that the physics is governed by a single parameter
${\X}=r_{\rm t}/r_{\rm l}$.  The same approximation formulas
(\ref{rt}) and (\ref{rl}) are used for the transverse ($r_{\rm t}$)
and longitudinal ($r_{\rm l}$) sizes, respectively, as it is done for
${\rm H}$-atom and ${\rm H}_2^+$. Their parameters are found by fitting the
results of calculations. The data for $r_{\rm t}, r_{\rm l}$ are
obtained using a strategy described in \cite{Turbiner:2004}. The
parameters of the fit are given in
Tables~\ref{table:6.1}-~\ref{table:6.2}. The fit of $r_{\rm t}$ and
$r_{\rm l}$ is illustrated in Figs.~\ref{fig:6.11}-~\ref{fig:6.12}.
Fig.~\ref{fig:6.13} demonstrates the behavior of the $\X$. The binding
energy $E_b$ which is calculated in \cite{Turbiner:2004} is
approximated using the formula (\ref{BE}) (see Table~\ref{table:6.3}
for parameters of the fit). The fit is illustrated in
Fig.~\ref{fig:6.14}.

In the same way as it is done for ${\rm H}_2^+$, we assume that the
equilibrium distance between protons are mostly defined by the
longitudinal size of the electron cloud (see (\ref{rl})), which are slightly
modified by including the terms depending on ${\X}_{\rm l}=\log {\X}$.
Finally, the equilibrium distance is approximated by Eq.(\ref{Req}) as was done for
${\rm H}_2^+$ (see Fig.~\ref{fig:6.12}). The parameters of the fit are given
in Table~\ref{table:6.4}. It is worth mentioning that the parameters
$c_i,\ i=0,1,2$ decrease very fast with $i$. This might be considered
as an indication of adequateness of the approximation formula
(\ref{Req}).

In the fit, some irregularities can be seen in the region $(5-50)
\times 10^{10}$\,G, near the threshold of appearance of the ${\rm H}_3^{++}$
ion (see Figs.~\ref{fig:6.11}-~\ref{fig:6.14}) similarly to those that
were observed for the \hbox{${\rm H}$-atom} and for the ${\rm H}_2^+$-ion.  One of
the reasons for these irregularities can be related to highly
increased technical difficulties we encountered exploring this region.
This could lead to a loss of accuracy. The overall quality of the fit
for the range $10^{11} - 4.414 \times 10^{13}$\,G is very high, 1-5\,\%.

Similarly to what was done for the ${\rm H}$ atom and the ${\rm H}_2^+$ molecular
ion, we calculate the expectation values $<|z|^n>\ ,\ n=2,3,4,5$ for
the ${\rm H}_3^{++}$ ion. It turns out that these expectation values admit a
very accurate polynomial approximation in terms of the expectation
value $<|z|>$, see Eq.(\ref{expect}). It seems natural to assume that
(\ref{expect}) holds for any $n$. The ground state eigenfunction
integrated over $\rho$ seems to define a certain one-parametric
distribution.  A similar phenomenon occurs for the ${\rm H}$ atom and the
${\rm H}_2^+$ molecular ion.

\section{The $({\rm HeH})^{++}$ molecular ion}

\begin{figure}
\begin{center}
   \includegraphics*[width=2.25in,angle=-90]{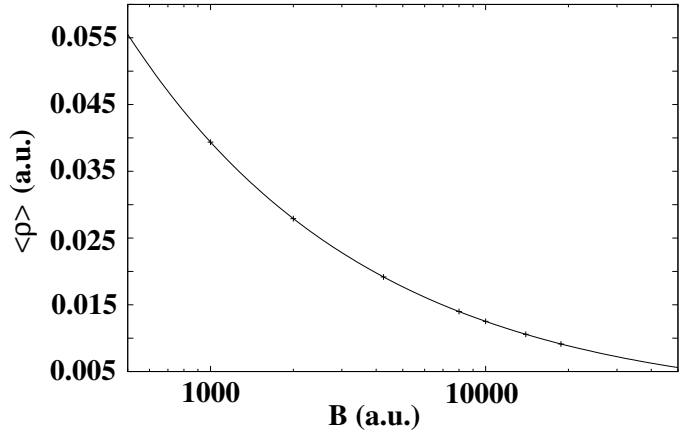}
    \caption{$({\rm HeH})^{++}$ ion: the fit of the  transverse size of the
    electron cloud $<\rho>$ using Eq.(\ref{rt}).
    Calculated values are indicated by crosses.}
   \label{fig:6.15}
\end{center}
\end{figure}

\begin{figure}
\begin{center}
   \includegraphics*[width=2.25in,angle=-90]{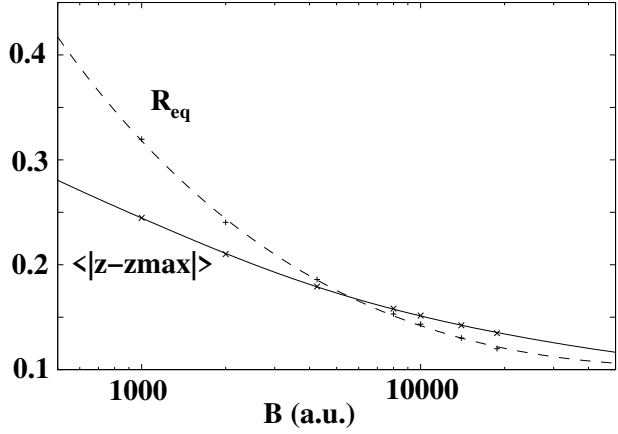}
    \caption{$({\rm HeH})^{++}$ ion: the fit of longitudinal size of the
      electron cloud $<(z-z_{\rm max})>$ using Eq.(\ref{rl}) (solid
      line) and the equilibrium distance $R_{\rm eq}$ using
      Eq.(\ref{Req}) (dashed line). Calculated values are indicated by
      crosses. All data in a.u.  }
   \label{fig:6.16}
\end{center}
\end{figure}

\begin{figure}
\begin{center}
   \includegraphics*[width=2.25in,angle=-90]{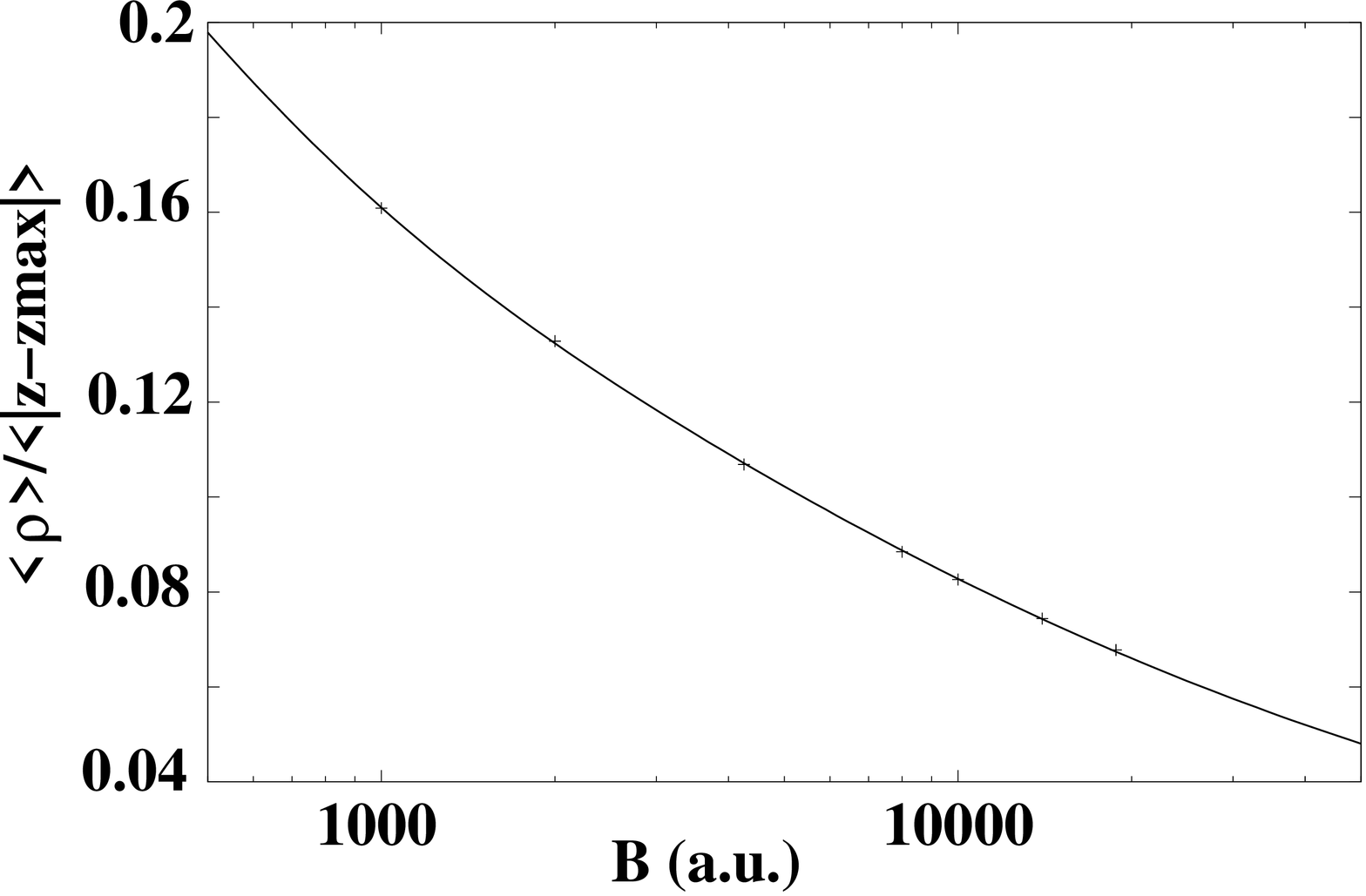}
    \caption{$({\rm HeH})^{++}$ ion: the ratio
    ${\X}=\frac{<\rho>}{<(z-z_{\rm max})>}$. Calculated values are
    indicated by crosses.}
   \label{fig:6.17}
\end{center}
\end{figure}

\begin{figure}
\begin{center}
   \includegraphics*[width=2.25in,angle=-90]{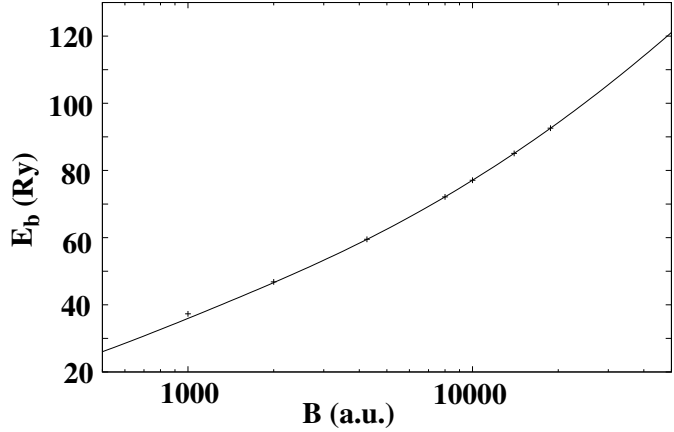}
    \caption{$({\rm HeH})^{++}$ ion: the fit of the binding energy
    using (\ref{BE}). Calculated values are indicated by crosses.}
   \label{fig:6.18}
\end{center}
\end{figure}

Recently, it was theoretically predicted that the exotic molecular ion
$({\rm HeH})^{++}$ can exist for $B \gtrsim 10^{12}$\,G
\cite{Turbiner:2004He}. Following the same idea of approximation as it
was implemented for the ${\rm H}$ atom and for the ${\rm H}_2^+, {\rm H}_3^{++}$
molecular ions, we can construct high-accuracy approximations for the
exotic $({\rm HeH})^{++}$ ion. Transversal $(r_{\rm t})$ and longitudinal
$(r_{\rm l})$ sizes~\footnote{The $({\rm HeH})^{++}$ molecular ion is
  characterized by the asymmetric electronic cloud. Therefore, the
  longitudinal size is defined $\nobreak{r_{\rm l} \equiv <(z-z_{\rm
      max})> \Big\vert_{z \ge z_{\rm max}}} - <(z-z_{\rm max})>
  \Big\vert_{z < z_{\rm max}}$ where $z_{\rm max}$ corresponds to the
  $z$-position of the maximum of the electronic distribution.}  of the
electron cloud as a function of the magnetic field are approximated by
the expressions (\ref{rt}) and (\ref{rl}) (see Figs.~\ref{fig:6.15}
and \ref{fig:6.16}). The parameters of the approximations
(\ref{rt})-(\ref{rl}) obtained through fitting the data from
\cite{Turbiner:2004He} are presented in
Tables~\ref{table:6.1}-\ref{table:6.2}. In Fig.~\ref{fig:6.17} the
ratio $\X$ is compared with the calculated data from
\cite{Turbiner:2004He}. The fit of the binding energy was performed
using the formula (\ref{BE}) (see Fig.~\ref{fig:6.18}).  The
parameters of the fit are presented in Table~\ref{table:6.3}.  For the
equilibrium distance $R_{\rm eq}$, the approximation (\ref{Req}) is
used (see Fig.~\ref{fig:6.16}); the parameters are presented in
Table~\ref{table:6.4}. The overall quality of the fit for the range
$10^{12} - 4.414 \times 10^{13}$\,G is very high, around 1\,\%.


\section{${\rm He}_2^{3+}$ molecular ion}

\begin{figure}
\begin{center}
   \includegraphics*[width=2.5in,angle=-90]{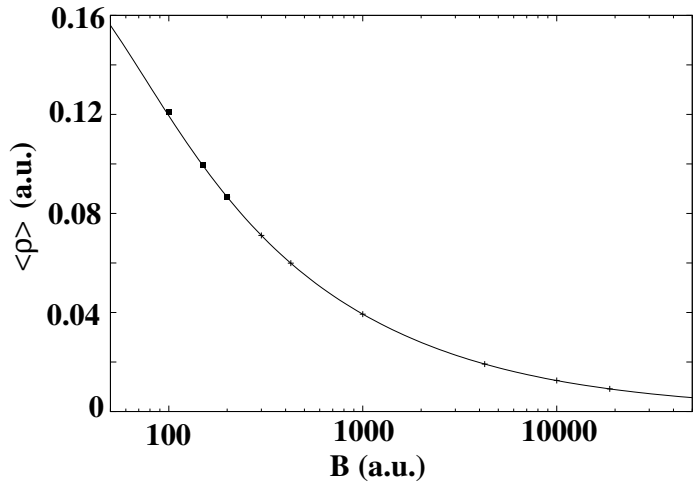}
    \caption{${\rm He}_2^{3+}$ ion: the fit of the  transverse size of the
      electron cloud $<\rho>$ using Eq.(\ref{rt}).  Only calculated
      values indicated by crosses, are used for fitting, while the
      calculated values shown as black squares are not taken into
      account (see text).}
   \label{fig:6.19}
\end{center}
\end{figure}

\begin{figure}
\begin{center}
   \includegraphics*[width=2.5in,angle=-90]{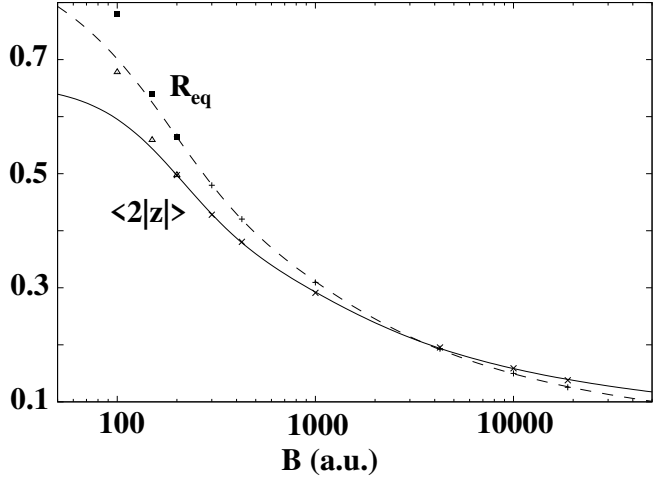}
    \caption{${\rm He}_2^{3+}$ ion: the fit of longitudinal size of the
      electronic cloud $2<|z|>$ using Eq.(\ref{rl}) (solid line) and
      of the equilibrium distance $R_{\rm eq}$ using Eq.(\ref{Req})
      (dashed line). All data in a.u.  Only calculated values which
      are indicated by crosses are used to make a fit, while
      calculated values shown by triangles (for longitudinal size
      data) and by black squares (equilibrium distance data) are not
      taken into account (see text).  }
   \label{fig:6.20}
\end{center}
\end{figure}

\begin{figure}
\begin{center}
   \includegraphics*[width=2.5in,angle=-90]{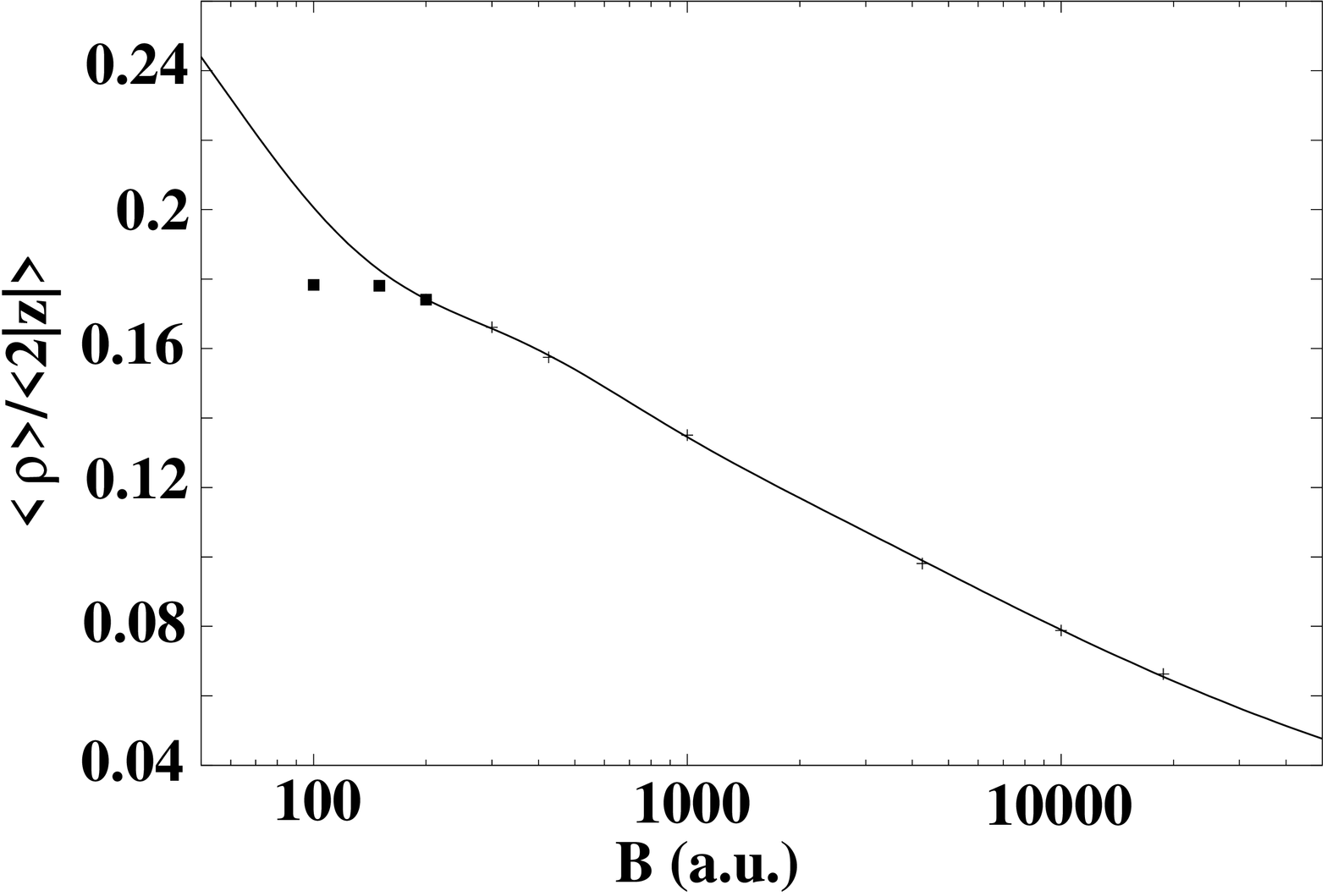}
    \caption{${\rm He}_2^{3+}$ ion: the ratio
    ${\cal X} = \frac{< \rho > }{ 2<|z|>}$.}
   \label{fig:6.21}
\end{center}
\end{figure}

\begin{figure}
\begin{center}
   \includegraphics*[width=2.5in,angle=-90]{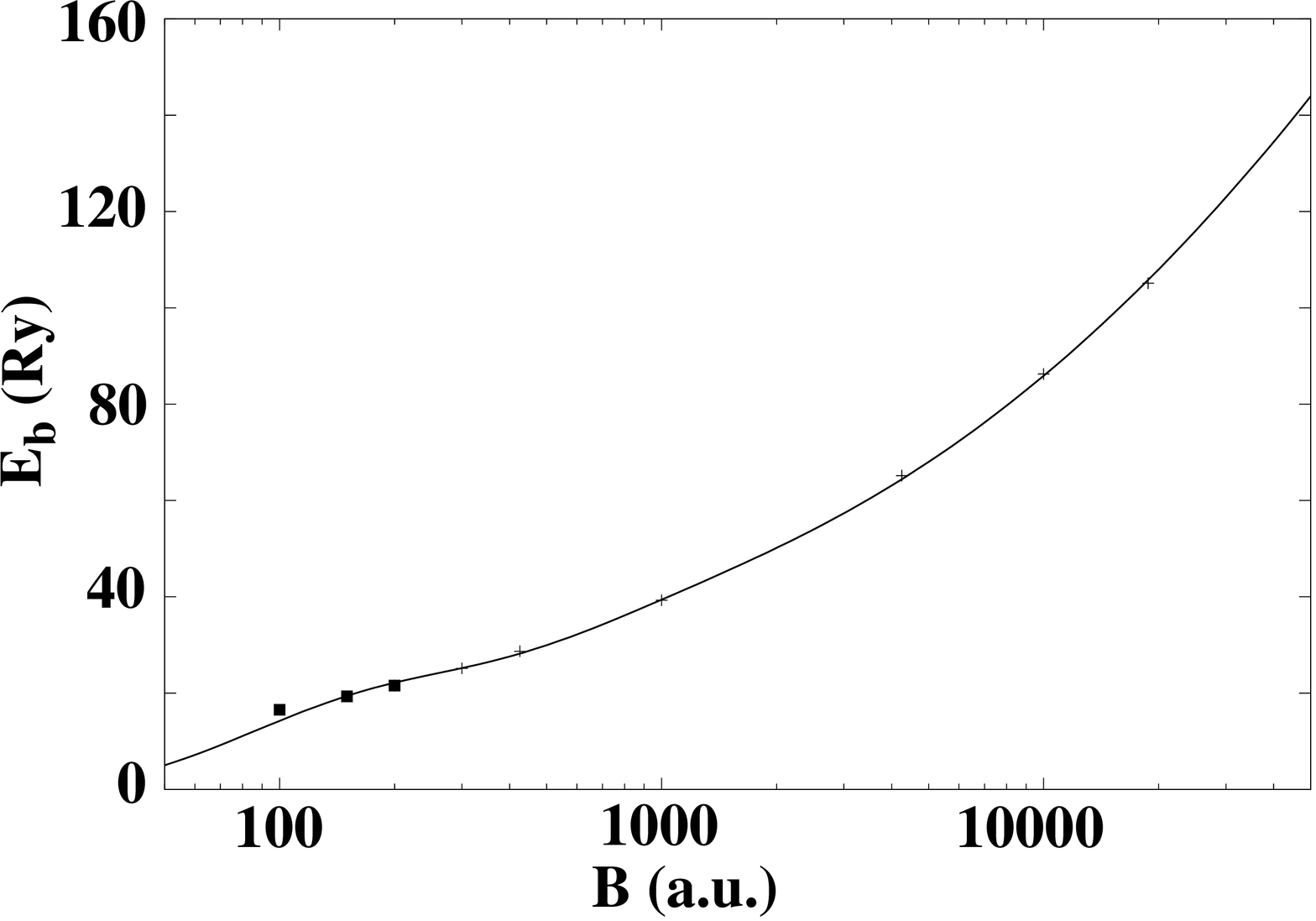}
    \caption{${\rm He}_2^{3+}$ ion: the fit of the binding energy
    using (\ref{BE}).}
   \label{fig:6.22}
\end{center}
\end{figure}

Recently, it was theoretically predicted that for $B \gtrsim 100\,$
a.u. the exotic molecular ion ${\rm He}_2^{3+}$ can exist
\cite{Turbiner:2004He}. Following the same idea of approximation as
for the ${\rm H}$ atom and the ${\rm H}_2^+, {\rm H}_3^{++}, ({\rm HeH})^{++}$ molecular ions
(see previous Sections), we would like to construct accurate
approximations for the exotic ${\rm He}_2^{3+}$ ion. Transversal $(r_{\rm
  t})$ and longitudinal $(r_{\rm l})$ sizes of the electron cloud as a
function of the magnetic field are approximated by the expressions
(\ref{rt}) and (\ref{rl}), respectively (see
Figs.~\ref{fig:6.19}~-~\ref{fig:6.20}). The parameters of the
approximations (\ref{rt}) and (\ref{rl}) obtained through the fit of
the data obtained in \cite{Turbiner:2004He} are presented in
Tables~\ref{table:6.1}-~\ref{table:6.2}, respectively. In
Fig.~\ref{fig:6.21} the ratio $\X$ is compared with the calculated
data from \cite{Turbiner:2004He}. The fit of the binding energy was
performed using the formula (\ref{BE}) (see Fig.~\ref{fig:6.22}). The
parameters of the fit are presented in Table~\ref{table:6.3}. For the
equilibrium distance $R_{\rm eq}$ the approximation (\ref{Req}) is
used (see Fig.~\ref{fig:6.20}) with parameters presented in
Table~\ref{table:6.4}.

Some irregularities can be seen in the fit in the region $(2-5) \times
10^{11}$\,G, near the threshold of appearance of the ${\rm He}_2^{3+}$ ion
(see Figs.~\ref{fig:6.19}-~\ref{fig:6.22}) similar to those which were
observed for the \hbox{${\rm H}$ atom} and for the ${\rm H}_2^+, {\rm H}_3^{++}$ ions.
The overall quality of the fit for the region $\nobreak{10^{12} -
  4.414 \times 10^{13}}$\,G is very high, around 1\,\%.


\newpage

\section{  {\it Conclusion}}

In this work we presented a phenomenological model of the behavior
of different one-electron atomic-molecular systems in a strong
magnetic field. The model is based on a surprisingly simple
physical idea that the ground state depends on a ratio of
transverse to longitudinal size of a system placed in a strong
magnetic field only. Since accurate numerical studies in a strong
magnetic field are very tedious from a technical point of view a
construction of a phenomenological model which provides
approximate expressions for basic characteristics of a system for
any value of a magnetic field strength can be quite useful for
applications.

One of the motivations of the present work is related to the fact that
the neutron star atmosphere is characterized by strong magnetic
fields, $10^{12} - 10^{13}$\,G. It seems natural to anticipate a
wealth of new physical phenomena there. However, for many years the
observational data did not indicate anything unusual, corresponding to
the black-body radiation. On 2002, the {\it CHANDRA} $X$-ray
observatory collected data on an isolated neutron star 1E1207.4-5209
which led to the discovery of two clearly-seen absorption features at
$\sim 0.7$\,keV and $\sim 1.4$\,keV \cite{Sanwal:2002}. It is
necessary to mention that the {\it XMM-Newton} $X$-ray observatory
recently confirmed the results of {\it Chandra/ACIS} related to
absorption features at 0.7 and 1.4 keV \cite{Bignami:2003}. We
proposed a model of hydrogen atmosphere with main abundance of the
exotic ${\rm H}_3^{++}$ molecular ion which explains these absorption
features assuming that the surface magnetic field is $\sim 5 \times
10^{14}$\,G \cite{Turbiner:2004m}.  For other neutron stars,
observational indications of the existence of absorption lines in
their spectra were already found\cite{Kerkwijk:2004,Vink:2004}. It
seems natural to anticipate forthcoming observations of other neutron
stars which will likely reveal absorption features. The study
presented here can be of certain use in identifying possible
absorption features.

\section{  {\it Acknowledgement}}

One of us (ABK) is grateful to the Instituto de Ciencias
Nucleares, UNAM, where the present work was initiated, for kind
hospitality extended to him. This work was supported in part by
CONACyT grant {\bf 36650-E} and DGAPA grant {\bf IN124202}
(Mexico), and by the RFBR grant 04-02-17263 and a grant of the
leading scientific schools 1774.2005.2 (Russia). AVT thanks the
University Program FENOMEC (UNAM) for partial financial support.

\end{document}